\documentclass[nofootinbib,
 reprint,
 amsmath,amssymb,
 aps,
]{revtex4-2}

\usepackage{graphicx}
\usepackage{dcolumn}
\usepackage{bm}
\usepackage{xcolor}
\usepackage{multirow}
\usepackage{hyperref}

\usepackage{babel}

\begin{document}

\title{Marr’s three levels for embryonic development:\\ information, dynamical systems, gene networks}

\author{David B. Br\"uckner}
\email{d.b.brueckner@unibas.ch}
\affiliation{
 Biozentrum and Department of Physics, University of Basel, Basel, Switzerland
}%

\author{Ga\v{s}per Tka\v{c}ik}
\email{gtkacik@ist.ac.at}
\affiliation{%
 Institute of Science and Technology Austria, Am Campus 1, AT-3400 Klosterneuburg, Austria
}%

\begin{abstract}
Developmental patterning comprises processes that range from purely instructed, where external signals specify cell fates, to fully self-organized, where spatial patterns emerge autonomously through cellular interactions. We propose that both extremes -- as well as the continuum of intermediate cases -- can be conceptualized as information processing systems, whose operation can be described using ``Marr's three levels of analysis'': the computational problem being solved, the algorithms employed, and their molecular implementation. At the first level, we argue that normative theories, such as information-theoretic optimization principles, provide a formalization of the computational problem. At the second level, we show how simplified information processing architectures provide a framework for developmental algorithms, which are formalized mathematically using dynamical systems theory. At the third level, the implementation of developmental algorithms is described by mechanistic biophysical and gene regulatory network models.
\end{abstract}

\maketitle

Development from early embryogenesis to organogenesis relies on the establishment of spatial patterns of distinct cell fates. These fates are induced by extracellular signaling factors, contact interactions, or mechanical cues that trigger intracellular cascades controlling gene expression dynamics~\cite{kicheva_control_2023,dupont_mechanical_2022}. During early development, patterns formed by such extra- and intracellular processes are typically classified into two broad categories~\cite{Collinet2021}. In the first category, initial symmetry breaking is instructed by external input signals. This paradigm is observed across diverse animal classes -- from worms and insects to amphibians -- exemplified by the early \emph{Drosophila} embryo, where maternally-produced morphogen gradients break initial symmetry and drive reproducible developmental outcomes through precise signal establishment and readout~\cite{gregor_probing_2007}. In the second category, symmetry breaking and subsequent patterning occurs autonomously through self-organization, proceeding in the complete absence of external spatial inputs. For instance, early mammalian embryos exemplify this paradigm: cells appear indistinguishable until the eight-cell stage, after which self-organized polarity establishment spontaneously occurs~\cite{Wennekamp2013}, possibly by amplification of small preexisting random inhomogeneities~\cite{lamba2024asynchronous}.

While instructed versus self-organized paradigms provide useful conceptual anchors, they are insufficient to capture the full richness of developmental processes. Most combine both paradigms. For example, the fly embryo initially patterns through maternal inputs but later exhibits self-organization guided by these initial inputs. Conversely, self-organized processes may incorporate modules where autonomously generated patterns provide inputs to subsequent instructed layers. Multiple layers of the developmental cascade may serve to refine or complexify the previous layer's pattern, or to make the system increasingly robust or invariant under system-size changes~\cite{barkai_variability_2007}. Whereas current approaches typically focus on a single layer that implements one of the prototypical patterning paradigms, we wish to conceptualize a richer, hierarchically-organized space of continuous patterning processes that combine and layer to orchestrate development. 

Yet how can we hope to navigate the space of such complexity? The ultimate results of development are diverse and adapted to specific external constraints, such as the development of paws, claws, wings or fins, depending on the ecological niche occupied by an animal. However, all developmental systems share the key feature that they generate patterns that are reproducible across embryos, which is fundamental for the development of a functional body plan, by ensuring the proper composition and thereby function of tissues in organogenesis. This precision is surprising given the inevitable presence of stochastic fluctuations at the cellular and subcellular scale, at which cell fate decisions are ultimately made: individual cells must ``measure'' extracellular input signals over time and make fate decisions based on these inputs. Due to inevitable biophysical constraints, such as the thermal or counting noise in molecule numbers, the signals received by single cells are subject to stochastic fluctuations, which limit the precision of cell fate decisions~\cite{balazsi_cellular_2011,raser_noise_2005}. Consequently, for spatially precise cell fate patterns to emerge, input signals must carry sufficient information, and individual cells must employ effective strategies to extract and process this information.

Many theoretical approaches have investigated the principles of cell-cell communication and cellular signal interpretation in the context of developmental patterning. These range from theories framed as optimization principles of information transmission through regulatory networks~\cite{tkavcik2025information,moghe_optimality_2025}, and information-theoretic measures of positional information~\cite{Dubuis2013,Tkacik2021a} or self-organization~\cite{Brueckner2024a}, to dynamical systems models of fate decisions~\cite{rand_geometry_2021}, as well as corresponding mechanistic frameworks such as reaction-diffusion and gene regulatory network models~\cite{morelli_computational_2012}. We propose that these approaches can be viewed as distinct yet complementary, by appealing to a perspective with a venerable history in neuroscience. Originally proposed by David Marr and Tomaso Poggio~\cite{marr1976understanding,marr_vision_2010}, any information processing system can be analyzed at three levels, now often referred to as ``Marr's three levels'': (1) the computational problem to be solved; (2) the algorithm employed; and (3) its implementation in the physical hardware available to the system. Here, we argue that optimization principles, which can be stated in terms of information-theoretic objectives, formalize the first level of description, i.e., the computational problem to be solved during development. We then examine how known patterning processes implement various algorithms for signal transformations, and how these algorithms can be productively analyzed by appealing to hierarchically stacked layers implementing information processing. We conceptualize how these transformations could underlie the spectrum of developmental processes from instructed to self-organized patterning and discuss how dynamical systems theory provides a unifying framework that formalizes these algorithmic transformations. Finally, we examine how the computational and algorithmic levels can guide us in analyzing and interpreting mechanistic models of developmental patterning.

\begin{figure}[t]
	\includegraphics[width=0.5\textwidth]{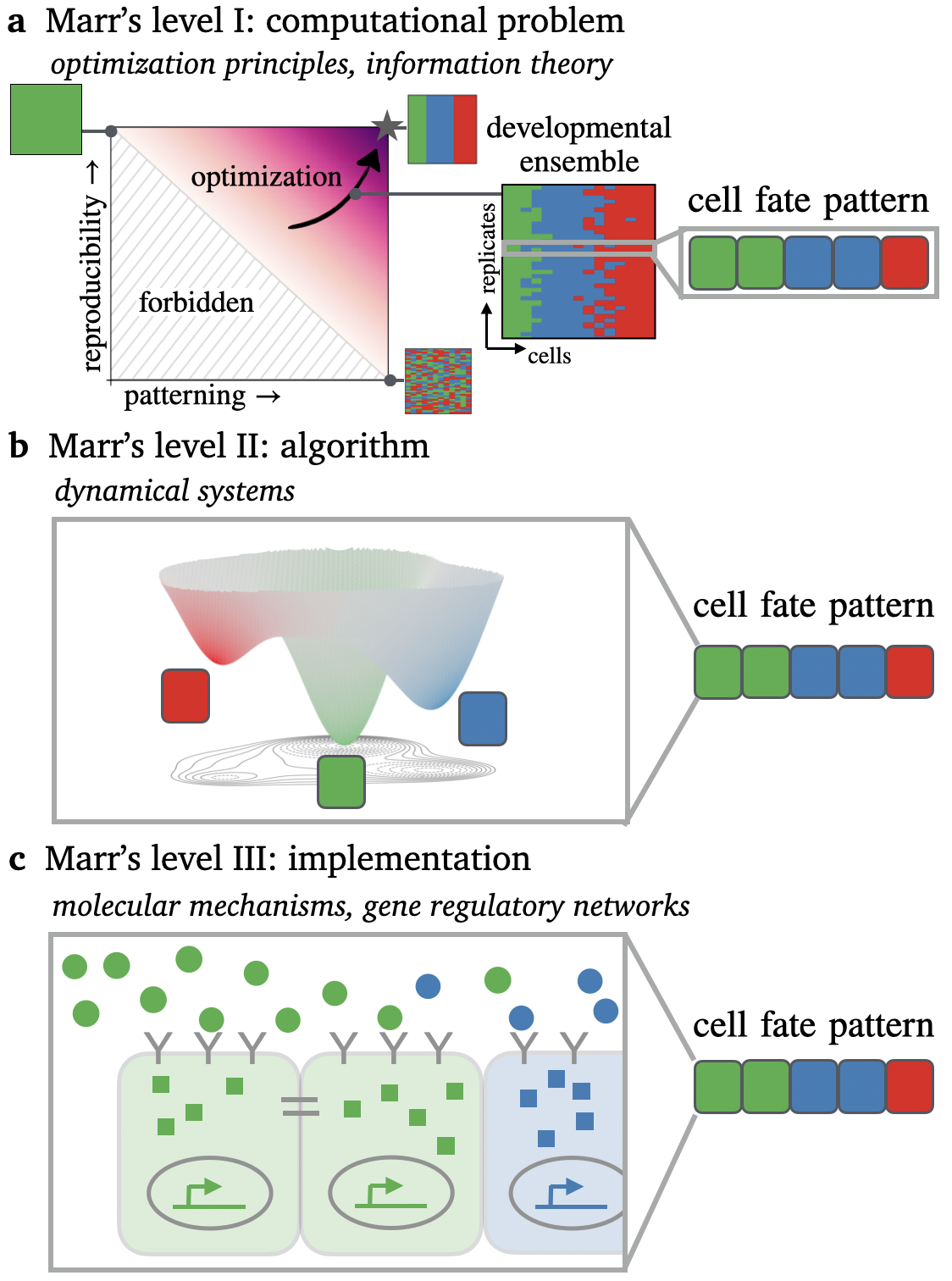}
	\centering
		\caption{Illustration of Marr's three levels for developmental biology.
        \textbf{a}, The computational problem can be defined using optimization principles such as maximizing information content, which drives patterns to high patterning entropy, high reproducibility states~\cite{Brueckner2024a}.
        \textbf{b}, The algorithmic level is formalized by dynamical systems theory modeling single and collective cellular states as valleys in a landscape~\cite{rand_geometry_2021}.
        \textbf{c}, The algorithms are implemented by molecular mechanisms and gene regulatory networks.
				 }
	\label{fig:levels}
\end{figure}

\section{Marr's first level:\\Normative theories for development}

The first level of Marr's analysis comprises a computational theory of a system by identifying what the system should compute and why. In the developmental context, this goal-oriented approach was pioneered early on: Wolpert's original article on the French Flag did not propose a specific ``model,'' but rather posed the problem of tissue patterning, to which multiple potential solutions were presented~\cite{wolpert_french_1968,Sharpe2019}. At Marr's first level, developmental problems are encapsulated by mathematical objective functions, sometimes also known as utility functions. Finding the best solutions to the considered developmental problem then corresponds to the optimization of these functions\footnote{Interestingly, this approach to biological systems shares many similarities with an influential approach to traditional sub-fields of physics, which also (re)state many of their fundamental laws as extremization or ``least-action'' principles.} (Fig.~\ref{fig:levels}a). In contrast to statistical and mechanistic models derived from experimental data or known underlying mechanisms, the normative approach starts by formulating a biologically motivated hypothesis that the system of interest likely evolved to perform a certain function. This description is subsequently made mathematically precise by writing down a corresponding objective function to be optimized. The focus of normative approaches thus remains firmly and constantly fixated on biological function: this focus -- as opposed to the symmetries and conservation laws governing our theories of inanimate matter --  is our guide to navigating biological complexity.

\begin{table*}
\caption{\label{table} Overview of theoretical approaches to developmental biology organized using Marr's three levels of analysis.}
\begin{ruledtabular}
\begin{tabular}{|c|c|p{12.5cm}|}
\textbf{Marr's levels} & \textbf{\begin{tabular}{c}Type of\\description\end{tabular}} & \begin{tabular}{c}\textbf{Examples}\end{tabular}  \\
\hline
{\begin{tabular}{c}Marr's level I:\\\textbf{computational}\\\textbf{problem}\end{tabular}} 
& \textit{\begin{tabular}{c}normative\\theories\end{tabular}} 
& \begin{tabular}{c}
maximizing information~\cite{sokolowski2025deriving,Brueckner2024a}\\
optimal control of temporal inputs~\cite{Pezzotta2022}
\end{tabular} \hspace{0.5cm}
\begin{tabular}{c}
optimal Bayesian decisions~\cite{Petkova2019}\\
morphogenetic action principle~\cite{cislo_morphogenetic_2023}
\end{tabular} \\
\hline
\multirow{2}{*}{\begin{tabular}{c}Marr's level II:\\\textbf{algorithm}\end{tabular}} 
& \textit{\begin{tabular}{c}algorithmic\\building blocks\end{tabular}} 
& \begin{tabular}{c}
thresholding~\cite{kracikova_threshold_2013}\\
temporal integration, filtering, adaptation~\cite{teague_time-integrated_2024,dessaud_interpretation_2007}
\end{tabular} \hspace{1cm}
\begin{tabular}{c}
spatial averaging~\cite{erdmann_role_2009}\\
lateral inhibition~\cite{Corson2017b}
\end{tabular} \\
\cline{2-3}
& \textit{\begin{tabular}{c} algorithmic\\models\end{tabular}} 
& \begin{tabular}{c}
French Flag Model~\cite{wolpert_french_1968}\\
Geometric Dynamical Systems Methodology~\cite{rand_geometry_2021} 
\end{tabular} \hspace{0.2cm}
\begin{tabular}{c}
Clock-and-Wavefront Model~\cite{cooke1976clock}\\
\end{tabular} \\
\hline
\multirow{2}{*}{\begin{tabular}{c}Marr's level III:\\\textbf{implementation}\end{tabular}} 
& \textit{\begin{tabular}{c}mechanistic\\building blocks\end{tabular}} 
& network motifs~\cite{alon_network_2007} \hspace{0.5cm} transcription factor activation \hspace{0.5cm} Delta-Notch signaling~\cite{gozlan_notch_2023} \\
\cline{2-3}
& \textit{\begin{tabular}{c}mechanistic\\models\end{tabular}} 
& \begin{tabular}{c}
reaction-diffusion \\ models~\cite{morelli_computational_2012}
\end{tabular} \hspace{1cm}
\begin{tabular}{c}
mechano-chemical \\
models~\cite{bailles_mechanochemical_2022}
\end{tabular} \hspace{1cm}
\begin{tabular}{c}
gene regulatory \\
network models~\cite{alon_introduction_2006}
\end{tabular} \\
\end{tabular}
\end{ruledtabular}
\end{table*}

The normative approach \emph{does not} assert that evolution has, in fact, driven developmental systems to their peak performance. Whether or not this occurred remains a separate empirical question, to be tackled on a case-by-case basis through collaborative experimental and theoretical work. A normative theory does, however, provide a mathematically precise hypothesis about how biological systems \emph{should} perform under fundamental physical and known biological constraints~\cite{smith1985developmental}. These quantitative hypotheses can be tested against experimental data directly, with significantly more (much needed!) nuance\footnote{For instance, the classic dichotomy of ``Did evolution push the system towards a theoretical optimum or not?'' is too narrow, polarizing, and often unproductive way  to think about evolutionary adaptation; being able to formulate the question as ``Is there empirical evidence that a particular function has been optimized by evolution?'' seems much more worthwhile.} as well as with statistical rigor, if desired~\cite{Mlynarski2021,zoller2022eukaryotic}. The conceptual distinction -- between asserting \emph{a priori} that evolution must have delivered a single optimal solution vs. deriving the optimal solution and treating the derivation as a quantitative prediction to be tested -- is absolutely essential.

Normative theory predictions that are quantitatively confirmed suggest that natural selection may have indeed pushed a biological system against fundamental physical limits. Substantial evidence for such optimization has accumulated in early vision~\cite{bialek2012biophysics}, organization of early sensory processing~\cite{balasubramanian2009receptive}, some biochemical signaling networks~\cite{tkavcik2025information}, and early development~\cite{sokolowski2025deriving}. Yet even when strict optimality for a single objective function is not achieved -- which is often the case -- or we are dealing with a hypothesized multi-objective evolutionary optimization~\cite{shoval2012evolutionary}, the comparison between optimal and evolved solutions can reveal shared organizational principles that can rationalize why particular biological solutions emerged from the evolutionary process~\cite{edwards2001silico,de2018statistical}. The normative frameworks we present (Table~\ref{table}) should therefore be viewed as providing precise, quantitative tools for understanding developmental design principles, rather than making sweeping claims about evolutionary history and adaptation.

Since all developmental systems start from a single fertilized cell, a central computational problem to be solved during development is to transform an aggregate of initially identical cells into a patterned array of distinct cell types in a manner that is minimally variable across an ensemble of embryos~\footnote{An interesting family of patterning problems (e.g., for Turing-type quasi-periodic spots) or spatial structures (e.g., in branching systems, or cortical connectivity) involves cases where reproducibility across embryos is not required on a cell-by-cell level, but on a statistical level, as captured by the correlational information formalism~\cite{Brueckner2024a}.}, despite stochastic fluctuations at all layers and across spatial scales. In early development, such reproducibility of non-trivial cell fate patterns at the cellular scale determines the emergence of a defined ``body plan'' at the collective level -- the evolutionarily-selected-for biological function of any multicellular developmental system. Therefore, to navigate the complex space of developmental processes, reproducibility in the presence of noise and environmental fluctuations needs to take center stage. The required language for that is statistical from the outset: we must think in terms of full probability distributions describing outcomes of a patterning process at the level of an ensemble of embryos, rather than a single individual or a simple mean over the ensemble. A normative theory applicable across diverse developmental systems would therefore score such reproducibility of the resulting ``body plans'' highly. Information-theoretic language is perfectly suited for the purpose: for instance, the spatial precision of a cell fate pattern is identified by its positional information -- the mutual information, in bits, between gene expression (or other cell fate markers) and cell position~\cite{Dubuis2013}. This function can be used to estimate the information content of morphogen gradients, revealing constraints under which cells make decisions~\cite{Tkacik2021a}. In the context of normative theories, it can be used as an objective function for optimization of algorithmic or biophysical models~\cite{morishita_optimal_2008,Francois2010,sokolowski2025deriving,iyer_cellular_2023}, which we will discuss below. Beyond morphogen-driven positional patterning, many systems undergo spontaneous symmetry breaking and self-organization, which can be optimized by maximizing the total self-organized information of a pattern which mathematically generalizes positional information~\cite{Brueckner2024a} (Fig.~\ref{fig:levels}a). While these information-theoretic quantities determine how much information is accessible to cells in principle, optimal Bayesian decision making has been used to optimize cellular strategies in inference problems, for instance in positional decoding~\cite{zagorski2017decoding,Petkova2019}, axon growth~\cite{Mortimer2009}, and intracellular decision making more generally~\cite{Libby2007,kobayashi_implementation_2010,Bowsher2014}. In addition to precision and reproducibility, the speed and metabolic costs of patterning have been formulated as objectives in an optimal control theory approach for the temporal dynamics of input signals driving cell fate decisions~\cite{Pezzotta2022}. Another approach used specific pattern features as an objective in evolutionary simulations, such as the number of segment boundaries~\cite{Francois2007a}, or the shape of a target pattern~\cite{booth_gene_2025}. Beyond cell fate patterning, normative approaches have also been proposed for morphogenesis, for instance in the formation of flow networks using flux optimization~\cite{ronellenfitsch_global_2016} or in three-dimensional tissue growth using action-like principles~\cite{cislo_morphogenetic_2023}. An exciting perspective for future work is to bridge the gap between pattern formation and morphogenesis, by merging them into a common normative framework.

\begin{figure}[b]
	\includegraphics[width=0.5\textwidth]{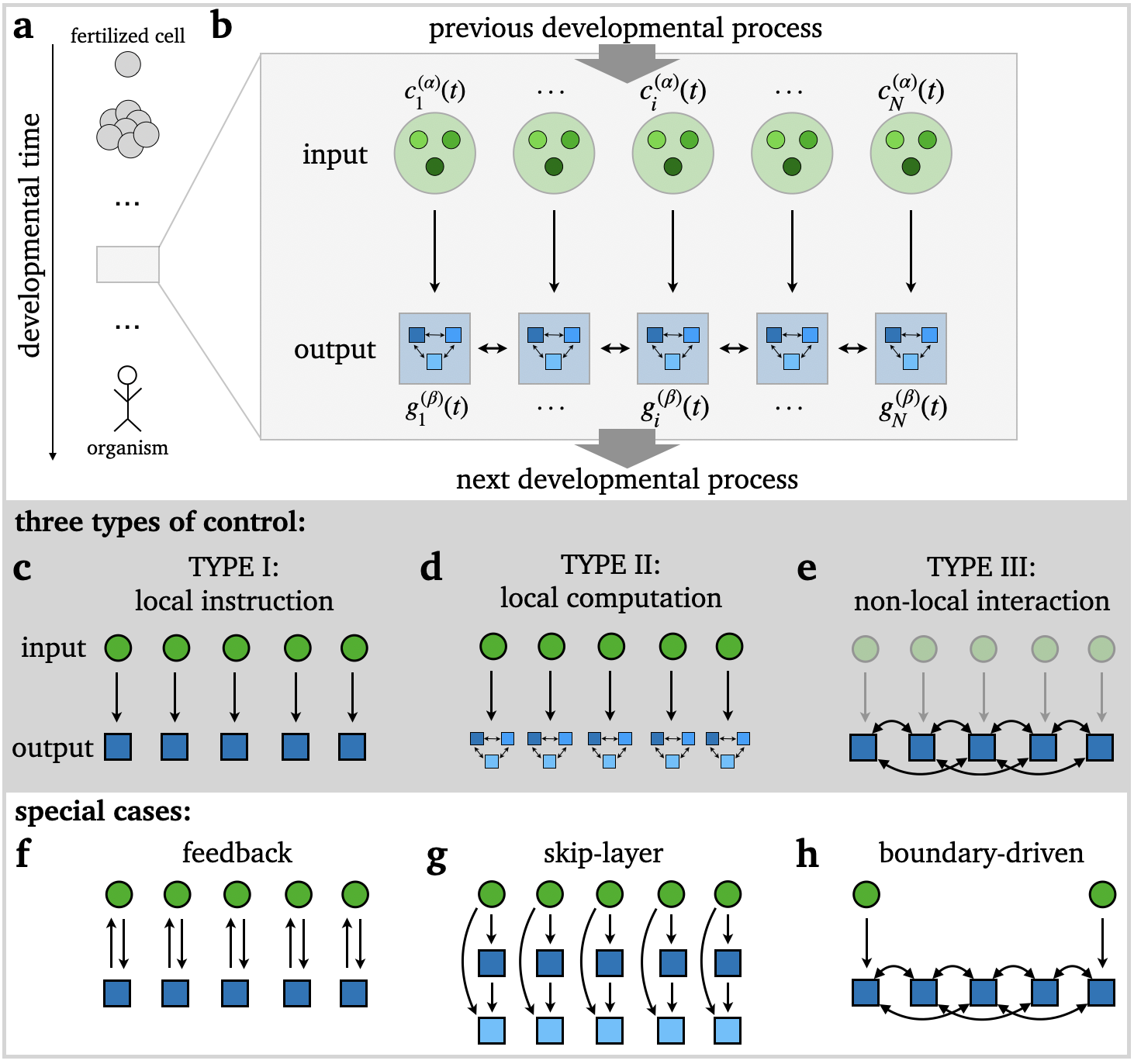}
	\centering
		\caption{Developmental information processing architectures.
        \textbf{a}, Schematic of developmental stages over time, from single fertilized cell to a fully-develped organism.
        \textbf{b}, A particular developmental process with fixed number of cells $N$ is represented as a two-layer system with inputs $c_i^{\alpha}(t)$ and outputs $g_i^{\beta}(t)$, with cell index $i$, various signals are denoted by indices $\alpha, \beta$, and time by $t$.
        \textbf{c}-\textbf{e}, Schematics of two-layer systems exhibiting three types of control. Special cases of such systems are feedback from outputs to inputs (\textbf{f}), skip-layer interactions (\textbf{g}), and boundary-driven systems (\textbf{h}).
				 }
	\label{fig:architectures}
\end{figure}

\section{Marr's second level: Developmental algorithms form an information-processing hierarchy}
Once the ``goal'' of the developmental system is formulated as an optimization problem on Marr's first level, one can look for algorithms that realize this goal or, more precisely, approach optimal solutions of the normative theory. This brings us to Marr's second level (Fig.~\ref{fig:levels}b). 

Here, we find it useful to distinguish between two categories of algorithms (Table~\ref{table}). On the one hand, we can consider basic operations or ``algorithmic building blocks'' that recur in various developmental settings (or, even more broadly, across cellular systems). Examples include thresholding operations via sharp nonlinearities~\cite{kracikova_threshold_2013}; gain amplification (ultrasensitivity)~\cite{zhang_ultrasensitive_2013}; linear or non-linear summation of signals; positive and negative feedback loops~\cite{alon_introduction_2006}; temporal integration~\cite{teague_time-integrated_2024}, filtering, and adaptation~\cite{dessaud_interpretation_2007}; as well as operations relying on spatial coupling including spatial averaging~\cite{erdmann_role_2009}, lateral inhibition~\cite{Corson2017b}, and coupled oscillations~\cite{doherty_coupled_2021}. On the other hand, we can consider fully fledged algorithmic models that solve a defined developmental problem and often utilize multiple algorithmic building blocks, while still describing the system without any direct reference to a particular mechanistic implementation. This category includes classic examples such as the French Flag Model~\cite{wolpert_french_1968}, Waddington's Landscape~\cite{waddington_strategy_1957}, the Clock-and-Wavefront Model~\cite{cooke1976clock}, as well as more recent Geometric Dynamical Systems Methodology~\cite{rand_geometry_2021}. 

The algorithmic level can be analyzed by decomposing each developmental process into an information-processing architecture that represents the tissue (Fig.~\ref{fig:architectures}a). Specifically, we consider a  set of discrete elements representing cells which may interact with one another, may be subject to externally provided (and potentially time-varying) inputs, and generate downstream outputs (Fig.~\ref{fig:architectures}b)~\cite{iyer_cellular_2023}.\footnote{For simplicity, we assume here a fixed number and spatial arrangement of cells. A major step forward in terms of normative theories and algorithms are extensions that simultaneously consider patterning along with cell proliferation, death, and spatial rearrangements.}
These information-processing architectures can be represented as a layered graph, in which the $N$ columns represent single cells, and rows represent successive layers of processing. The first layer represents external inputs $c_i(t)$ received by each cell $i=1,\dots,N$ as a function of time $t$. The inputs are processed to generate outputs $g_i(t)$ in the second layer. This description readily generalizes to multidimensional signals, represented as input and output vectors $\{c^{(\alpha)}_i(t)\}$ and $\{g^{(\beta)}_i(t)\}$, with greek letters indexing over vector components. At the algorithmic level, these components might -- but need not -- correspond to individual chemical species at the mechanistic level. For example, the first input layer could be (one or multiple) extracellular diffusible ligands, which regulate the expression of (one or multiple) downstream output genes.

How does the information-processing architecture, represented as a layered graph, relate to algorithms? Much like ``block diagrams'' that are essential to systems design in hardware, software, control, and process engineering, our layered graph partitions the developmental system by constraining which signals can interact and which cannot, with arrows corresponding to various algorithmic operations, as enumerated above, that act on the signals. We emphasize that our architectures differ from past uses of network diagrams in systems and developmental biology which typically represented intracellular gene regulatory networks, with blocks and arrows representing individual genes and their interactions~\cite{davidson2003regulatory,alon_introduction_2006,alon_network_2007}.
In contrast, the architectures explicitly reference space (columns), and assume a time-scale separation that allows the decomposition of the developmental trajectory into subprocesses with inputs and outputs, i.e., they directly reference time -- albeit in a coarse-grained fashion (rows). Within each layer, multiple species may locally interact (e.g., within a node of the graph, Fig.~\ref{fig:architectures}b), as realized mechanistically by chemical reaction or gene regulatory networks. Importantly, the classification of any signal as either input ($c$) or output ($g$) is inherently contextual rather than absolute. For example, during the early patterning of the fly embryo, we could identify the input signal $c$ with maternal gradients spanning the anterior-posterior axis, which drive the output $g$, the expression of central gap genes. However, in the subsequent layer of patterning, these same gap genes could be considered as inputs ($c$) controlling the expression of output pair-rule genes ($g$).

It is instructive to start by considering simple architectures. Among two-layer systems with local instruction we distinguish three types of control over output signals. The first is local instruction in which each input $c_i(t)$ is processed solely within cell $i$, meaning that the corresponding output $g_i(t)$ is a function of only $c_i(t)$ (Fig.~\ref{fig:architectures}c) \cite{tkacik_optimizing_2009}. Such scalar-to-scalar signal transformations can already realize powerful algorithmic operations: $g_i(t)$ can temporally process $c_i(t)$ by taking its average, compute its temporal derivative, perform frequency filtering more generally~\cite{de_ronde_feed-forward_2012}, or even detect or discriminate between temporal features in the input. Alternatively, $g_i$ could be a discrete marker gene that needs to be turned ON or OFF at a ``readout time'' $T$, depending on the entire temporal history of $c_i(t)$ for $t<T$~\cite{tkavcik2025information}. 

The second type of control occurs via multi-dimensional yet still local operations within each cell: intracellular responses to external signals can be the result of complex local computations due to gene regulatory networks whose interactions can give rise to new phenomenology (Fig.~\ref{fig:architectures}d) \cite{walczak_optimizing_2010,Tkacik2012a}. In contrast to scalar-to-scalar transformations characteristic for the first type of control, here the inputs drive multi-dimensional dynamical systems that can implement, for instance,  attractor landscapes, multistability~\cite{rand_geometry_2021}, or limit cycle oscillations. This opens up a much wider and, importantly, controllable richness of possible output behaviors, with the inputs plausibly regulating the bifurcations and operating regimes of dynamical systems running within individual cells~\cite{kadiyala2025genes}.

The third type of control permits non-local (spatial) interactions between the cells which are now allowed to exchange information (Fig.~\ref{fig:architectures}e) \cite{Sokolowski2015}. We focus on interactions at the level of outputs $g_i^{(\beta)}(t)$, since these interactions can exist even when there are no inputs, such as in self-organized development.\footnote{Inputs that interact \emph{before} being processed by the cells clearly shape what individual cells in the tissue would perceive, but this can be captured by the statistical structure (correlations) in the input.} This architecture corresponds to a processing system with ``lateral'' interactions: these can still be strictly feed-forward (depending on the interaction topology) or recurrent, but they do not feed back onto the inputs. Such interactions can be either short-ranged (e.g., direct nearest-neighbor interactions) or long-ranged (e.g., paracrine or via secretion of diffusible ligands). These  interactions couple what would otherwise be independent dynamical systems running in individual cells into a collective dynamical system, whose attractors -- now defined at a tissue, not cellular, level -- may lead to the establishment of spatial structure and reproducibility beyond the information that has been encoded in the input alone~\cite{hillenbrand2016beyond,Brueckner2024a}.

This decomposition into layers can provide a good approximation to a given developmental system for multiple reasons. One could empirically verify the time-scale separation between different layers in the system of interest, or confirm that developmental processes follow a strict temporal sequence; or establish that the information flow is predominantly feed-forward across developmental layers. Within this framework, multiple special cases may arise. First, inputs provided by the upstream developmental subprocesses may trigger feedback interactions from the downstream outputs back onto the inputs (Fig.~\ref{fig:architectures}f). For example, in activator-inhibitor systems such as Nodal-Lefty in zebrafish, Nodal is first produced, then activates Lefty, which in turn inhibits Nodal~\cite{Muller2012}. Second, when hierarchically stacking layers, the outputs of each layer may drive inputs to subsequent layers so as to allow ``skip-layer'' interactions (Fig.~\ref{fig:architectures}g). For example, in the fly embryo, enhancers for the pair-rule even-skipped gene also harbor bicoid binding sites, implying that some information might feed into the pair-rule pattern directly from maternal morphogens, skipping the gap gene layer~\cite{arnosti1996eve}. Importantly, while both of these cases have much in common with the second type of control through local computations (Fig.~\ref{fig:architectures}d), they may differ in phenomenology depending on the context; for instance if the different interactions operate on different time-scales. Finally, a special case of non-local interactions are boundary-driven systems: here, input is provided only at the system boundaries, thereby guiding internal interactions to self-organize the cells into a reproducible pattern. For example, 2D gastruloids follow this boundary-instructed route implemented by BMP signaling~\cite{Etoc2016}. Taken together,  information processing architectures, the simplest of which we highlighted here, provide an organizing scheme for signal processing operations implemented by various algorithmic building blocks (Fig.~\ref{fig:levels}), and may also provide a broader framework for the analysis of algorithmic models. 

One of the most promising research directions at the level of algorithmic models is the recent work on the Gene-Free or Geometric Methodology for cell fate decisions~\cite{corson2017gene,rand_geometry_2021} (Fig.~\ref{fig:levels}b). This methodology offers a broadly applicable mathematical framework that can implement various architectures discussed above, by representing developmental processes as flows through landscapes of attractors reshaped by bifurcations. It offers an elegant connection to an established and wide body of biology literature, by mathematically formalizing  Waddington's original metaphor. The resulting dynamics describe developmental decisions by Morse-Smale systems which admit potential functions decreasing along trajectories, with critical points corresponding to cellular decision states~\cite{rand_geometry_2021}. Crucially, such systems can be represented, without loss of generality, by minimal parameterizations that focus on the essential geometric features -- saddle points, unstable manifolds, and bifurcation structures -- rather than by tracking individual molecular species. The combination of mathematical rigor and theoretical guarantees~\cite{rand_geometry_2021} with the simplicity that can be related to population~\cite{Corson2017b} or single-cell data~\cite{Camacho-Aguilar2021,saez2022statistically,hajji_quantitative_2024,cislo_reconstructing_2025,fontaine_dynamic_2025}, makes this a very powerful and promising avenue of research.

How should we think about these approaches in the light of Marr's levels of analysis? One suggestion would be to view the information-theoretic formulation at Marr's first level as defining an optimization over dynamical systems landscapes of the Geometric Methodology so as to achieve reproducible body plans in face of internal and external sources of stochasticity. The Geometric Methodology at Marr's second level further suggests that while the molecular details may vary dramatically across species and contexts (Marr's third level), the fundamental algorithms underlying developmental information processing could exhibit convergent dynamical structures characterized by particular arrangements of attractors, saddles, and their connecting manifolds~\cite{saez_dynamical_2022,gallo_versatile_2024}. It also informs us about the relevant collective dynamical variables and control parameters shaping the geometric landscape, thus providing a clear guide to the molecular and mechanistic complexity that we discuss next.

\section{Marr's third level:\\Mechanistic implementation of developmental algorithms}
The developmental algorithms described in the previous section are implemented at the molecular and cellular scale through biochemical and biophysical mechanisms (Fig.~\ref{fig:levels}c). 

We again propose to partition these mechanisms into two categories (Table~\ref{table}). On the one hand, we can study individual mechanistic building blocks that have been evolutionarily reused across systems, such as network motifs, transcription factor activation and translocation mechanisms, controlled degradation, chemoreception mechanisms along with receptor internalization and adaptation, as well as cell-cell communication systems, such as the Delta-Notch pathway. On the other hand, we can construct full-scale mechanistic models that combine multiple such building blocks, such as reaction-diffusion~\cite{morelli_computational_2012}, mechanical~\cite{schwayer_connecting_2023}, mechano-chemical ~\cite{bailles_mechanochemical_2022,Brueckner2024b}, and gene regulatory network models~\cite{alon_introduction_2006}. These are mathematically specified with sets of ordinary differential equations for chemical reaction systems or partial differential equations for spatial (diffusion) processes; stochastic differential equations, stochastic simulations of chemical reactions systems, or agent-based simulations can be used where intrinsic stochasticity is important. Mechanistic models are typically instantiated and fitted to particular biological processes, for example for the early biochemical patterning in the fruit fly~\cite{sokolowski2025deriving,von2000segment}, zebrafish gastrula~\cite{Muller2012,pomreinke_dynamics_2017}, or vertebrate systems~\cite{sick_wnt_2006,sheth_hox_2012,raspopovic_digit_2014}; for mechanical morphogenesis~\cite{serra_mechanochemical_2023,caldarelli_self-organized_2024}; as well as self-organization of various organoid systems~\cite{Chhabra2019a,lehr_self-organized_2024,schwayer_cell_2025}. This modeling approach has the benefit of connecting most directly to experiments, especially if the model is able to predict the outcomes of novel interventions (genetic perturbations, acute or pharmacological perturbations, changed external conditions etc.). As such, its value -- more so than its pitfalls \emph{vis-a-vis} top-down models -- has been recognized by the biological community and widely reviewed; we therefore restrict our comments here only on how such models interact within our broader perspective.

We argue that especially for mechanistically-detailed models, typically very complex as they are constructed bottom up from a long list of known necessary biological components and mechanisms, the algorithmic and normative perspectives provide a powerful guide and an interpretative framework. For example, they allow us to reason about how the three algorithmic levels of control~(Fig.~\ref{fig:architectures}c-e) can be realized in molecular hardware. The first type of control can be implemented through transduction of an extracellular input signal $c$ via chemoreception into a cell-intrinsic representation, such as phosphorylation of a transducer molecule that drives gene expression $g$. The second type of control can be implemented by multiple such transducers engaging in intracellular crosstalk to preprocess the signals, or through the interactions of multiple cross-regulating genes downstream of the single signal transduction pathway. For example, receptor-ligand promiscuity in the BMP pathway allows cells to perform complex computations on combinations of different BMP ligands~\cite{antebi_combinatorial_2017}. At the gene level, the interpretation of Sonic Hedgehog dynamics by the downstream transcriptional network of mutually cross-repressing transcription factors Olig2, Pax6 and Nkx2.2 in neural tube patterning exemplifies this type of control~\cite{balaskas_gene_2012}. Spatial couplings in the third type of control could be implemented, for instance, through para- and juxtacrine signaling, or mechanical interactions between cells. Classical examples are juxtacrine signaling through the Delta-Notch pathway which can establish long-range order through local interactions~\cite{gozlan_notch_2023}, diffusive morphogens and their inhibitors which implement Turing patterning of hair follicles~\cite{sick_wnt_2006,Kondo2010}, as well as mechanical interactions that link to cell fate through mechanotransduction~\cite{dupont_mechanical_2022}.

Taken together, we argue that to build an overarching theory of embryonic development, we must merge these mechanistic approaches with the second and first of Marr's levels. By defining the computational problem, we can identify optimal algorithms, and subsequently determine how known mechanisms may implement these optimal algorithms in specific systems. In the next section, we explore how two limiting cases of these architectures at Marr's second level provide conceptual foundations for instructed versus self-organized development, and how these two architectures solve their corresponding normative problems at Marr's first level.

\begin{figure*}[t]
	\includegraphics[width=\textwidth]{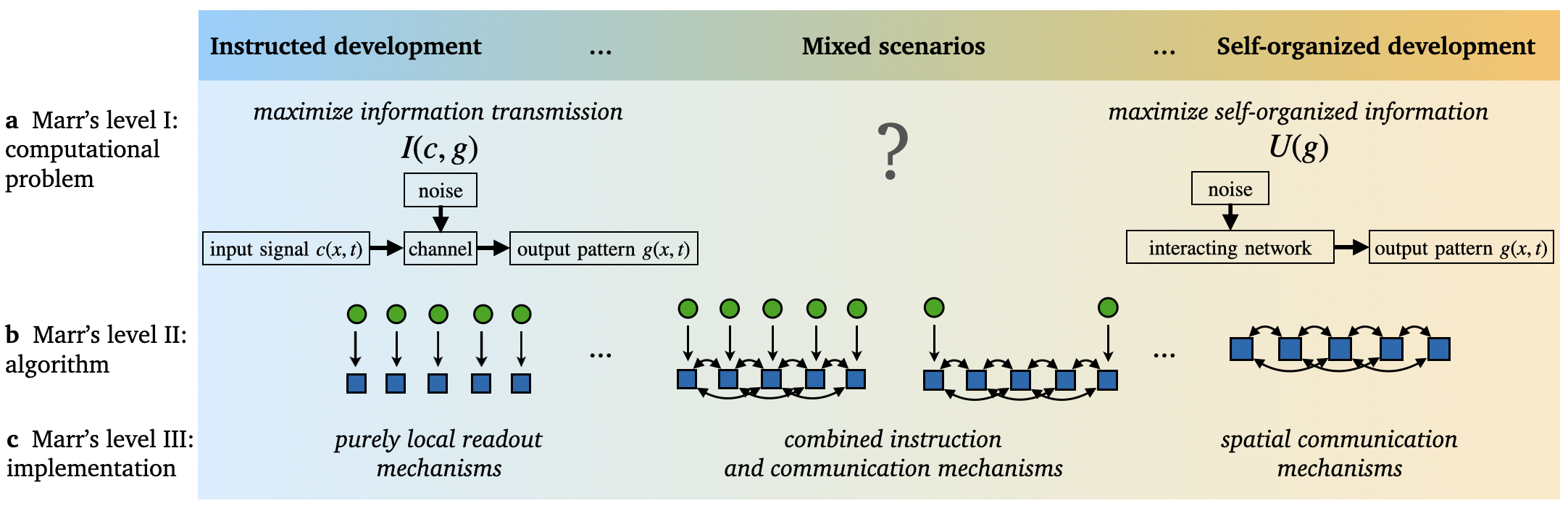}
	\centering
		\caption{Analyzing instructed and self-organized development across Marr's three levels.
        \textbf{a}, While instructed development can be formalized as an information transmission problem, the problem of self-organized development can be represented as transforming noise into an output pattern that maximizes self-organized information. Mixed scenarios have not yet been formalized at an information-theoretic level.
        \textbf{b}, The algorithmic architectures can be arranged along the axis from instructed to self-organized cases along decreasing importance of the input and increasing importance of spatial communication.
        \textbf{c}, These architectures are implemented by mechanisms that are purely local in instructed scenarios, but allow spatial communication in self-organized scenarios.
				 }
	\label{fig:spectrum}
\end{figure*}

\section{Instructed and self-organized development}

To illustrate how our framework applies to biological pattern formation, we focus on the two limiting cases of instructed and self-organized patterning, viewed across  Marr's three levels of analysis (Fig.~\ref{fig:spectrum}).

\textbf{Instructed development.} Algorithmic architectures where cells receive external input signals $c(x,t)$ that depend on position $x$ and time $t$ can be understood as formalizing the extreme limit of purely instructed patterning. Here, the processing of the input signals occurs locally and independently in each cell, without any lateral coupling or cell-cell coordination, thus corresponding to the first two types of control at the algorithmic level (Fig.~\ref{fig:spectrum}b). At Marr's first level, the computational problem corresponding to pure instructed development is to maximize information transmission from input signals to output states, construed as expression levels or, ultimately, cell fates (Fig.~\ref{fig:spectrum}a)~\cite{tkacik_optimizing_2009,walczak_optimizing_2010,Tkacik2012a}. In principle, this implies that the spatial precision of the output states $g(x,t=T)$ (at some final ``readout time'' $T$), quantified as the positional information $I(g,x)$, is constrained by the positional information of the spatially structured input (e.g., a set of morphogenetic gradients), $I(c,x)$, or the information transmission from inputs to outputs via the chemical reaction network operating within each cell, i.e., the mutual information $I(c,g)$, via one of the fundamental theorems of information theory, the Data Processing Inequality~\cite{cover1999elements}. Care needs to be taken in the interpretation of this inequality: since $g$ can typically temporally process the inputs $c$ -- via averaging, nonlinear transformations, thresholding etc., as implemented on Marr's third level by signaling cascades, post-translational modification, molecule translocation and so on -- the mutual information terms $I$ need to be defined over full signal time-trajectories, i.e., $c= \{c(0),...,c(t),...,c(T)\}$. Data processing inequality (DPI) then asserts that $I(g;x) \leq I(c;x)$ and $I(g;x) \leq I(g;c)$, where $g$ and $c$ are understood as full temporal trajectories of input and output signals, respectively (not instantaneous values!). The corresponding normative theory at Marr's first level then optimizes either the information transmission $I(g;c)$ if $c(x,t)$ is given and held fixed by experimental measurements, or the full ``positional information'' $I(g;x)$ directly, if the the morphogen gradient itself can also be considered to be an optimization target, with constraints on its noise and/or reproducibility. In either case, this approach quantitatively captures the essence of the classical positional information concept: cells can only infer their spatial positions from local morphogen measurements.

\textbf{Self-organized development.} In pure self-organization, there are no meaningful, spatially-structured external inputs $c(x,t)$. Instead, patterns emerge through spatial coupling between cells, thus corresponding to the third type of control at the algorithmic level (Fig.~\ref{fig:spectrum}b). At Marr's first level, the computational problem can therefore no longer be defined as maximizing the mutual information between the input and output signals, as in instructed development. Instead, a possible objective of the system is to turn unstructured noisy inputs into spatially patterned outputs that are as reproducible as possible (Fig.~\ref{fig:spectrum}a). Mathematically, this can be formulated as a utility function $U(g)$ that maximizes the total information content of the final output pattern $g(x,t=T)$ at some final time $T$, without any reference to the inputs~\cite{Brueckner2024a}. Such final state information can be compared to the amount of information exchanged laterally between cells, and recent work demonstrates that these two quantities are not necessarily maximized simultaneously~\cite{tripathi_collective_2025}. This suggests that optimal strategies at a collective level, determined by optimizing multicellular objectives, do not necessarily correspond to optimal behavior of individual cells when considered in isolation. At Marr's second level, collective objective functions enable a broader search through algorithmic rules that spatially couple cells, such as cellular automata~\cite{ermentrout_cellular_1993}, or collective dynamical systems which are defined without direct reference to underlying mechanisms~\cite{Corson2017b,rand_geometry_2021}. To enable self-organization, these algorithms must contain spatial couplings (Fig.~\ref{fig:architectures}e) which must ultimately be realized through specific biological mechanisms at Marr's third level, for example, as reaction-diffusion systems~\cite{morelli_computational_2012,kicheva_control_2023}, lateral inhibition signaling~\cite{Corson2017b}, or mechano-chemical circuits~\cite{bailles_mechanochemical_2022}.

\textbf{Mixed scenarios.} Most systems are probably neither purely instructed nor self-organized; they likely correspond to mixed scenarios combining elements of both extreme limits~\cite{green_positional_2015}. In this case, all three types of algorithmic control combine (Fig.~\ref{fig:spectrum}b). Evolved systems likely make use of external inputs to break the symmetry while strongly relying on downstream spatial cell-to-cell couplings. Such hybrid systems can be conceptualized in two ways. First, \emph{spatially-coupled instructed systems} receive ambiguous input signals which they spatial filter and refine so as to increase the information content of the final pattern. For example, noisy external gradients may undergo additional algorithmic operations, such as smoothing through spatial averaging~\cite{erdmann_role_2009}, divisive normalization~\cite{Brueckner2024a}, or sharpening through lateral inhibition~\cite{piddini_interpretation_2009}. Such nonlocal coupling mechanisms can enhance pattern precision beyond what pure information transmission could achieve in a strictly local setting. Second, \emph{instructed self-organizing systems} receive very rough, low-information-content external inputs that simply bias or ``kick'' the multi-stable system towards a particular attractor, thereby choosing among multiple possible (and possibly spatially complicated) collective output states. In this scenario, external signals do not directly specify the final pattern but rather guide the selection of self-organized outcomes. A key theoretical challenge for the future lies in unifying these two perspectives: by interpolating how systems transition between instruction-dominated and self-organization-dominated regimes, and developing the required mathematics to connect these limiting cases within a single theoretical framework.

\section{Open questions}

While the theoretical frameworks we have outlined provide a foundation for the understanding of developmental patterning, several fundamental questions remain open. They will be crucial for advancing this field toward a comprehensive theory of development.

\textbf{Multilayered systems.} While we focused our discussion on simplified two-layer input-output information processing systems, real developmental networks are inherently multilayered, often featuring complex feedback or skip-layer connections (Fig.~\ref{fig:architectures}). However, many models in developmental biology employ reduced descriptions that focus on a particular two-layer step of the entire cascade. At Marr's first level, this raises the question of how optima of single steps of the developmental hierarchy are distinct from ``end-to-end'' optimization of the whole hierarchy. Such end-to-end optimization raises the possibility of successive layers being ``matched'' to one another\footnote{This is perhaps reminiscent of, e.g., impedance matching in electrical engineering, or Kirchoff's laws in flow networks.} to optimize the final outcome in a way that is distinct from level-by-level optimization. At Marr's second level, this involves understanding how information flows and transforms across multiple processing stages, while at the third level, it requires characterizing how molecular networks can be organized to achieve optimal multilayer information transmission or processing.

\textbf{Robustness against natural fluctuations}. A crucial challenge for normative theory at Marr's first level lies in identifying the natural space of intrinsic fluctuations, environmental variations, and embryo-to-embryo variability, to which developmental systems must be robust. For instance, natural temperature fluctuations may have driven the evolution of specific regulatory motifs that would not be predicted if we only considered laboratory-controlled conditions~\cite{kuntz_drosophila_2014,mirth_growing_2021}. The concept of robustness against fluctuations in particular external variables is naturally described by the framework of reproducibility: robustness means reproducibility under the constraint that the considered external variable is fluctuating within its natural range. Just as neuroscience has benefited from distinguishing between laboratory-controlled perturbations and natural sensory statistics~\cite{simoncelli2001natural}, normative theories of development -- to be truly predictive -- will require a careful characterization of the ecological and environmental fluctuation patterns that have shaped the evolved developmental systems.

\textbf{Morphogenesis, growth and scaling}. The fixed-lattice, one-dimensional information processing architectures (Fig.~\ref{fig:architectures}) provide a first approximation for a static tissue, which will likely break in systems where patterning occurs simultaneously with growth, morphogenetic flows, three-dimensional shape changes, and active cellular motion~\cite{bailles_mechanochemical_2022,Brueckner2024b,plum_morphogen_2025}. While models of tissue mechanics have traditionally been formulated with direct reference to molecular mechanisms~\cite{Brueckner2024b}, recent papers have employed the geometrical approach to tissue mechanics, providing algorithmic descriptions of mechanical couplings and pattern formation at Marr's second level. Example work identified negative and positive mechanical feedback as ``algorithms'' responsible for mechanical stability~\cite{noll_active_2017} and self-organized convergent extension~\cite{brauns_geometric_2024}. Furthermore, ``Dynamic Morphoskeletons'' have been proposed as a dynamical systems formulation of morphogenetic modules, in analogy to the Geometrical Methodology for cell fate decisions~\cite{plum_dynamical_2025,serrano_najera_control_2024}. Beyond purely mechanical models, integrating the cell fate pattern with tissue deformations introduces additional layers of complexity: information must be preserved or appropriately transformed as the physical arrangement of cells changes. For instance, morphogen gradients in growing tissues often exhibit scaling properties that maintain biological function across different developmental stages and organism sizes~\cite{Ben-Zvi2008,Gregor2005,nikolic_scale_2023}. At Marr's first level, this is an example of robustness that can be thought of reproducibility under the constraint of natural fluctuations -- or indeed directed growth -- in embryo length. An open question is whether optimization under such constraints would identify new principles of system size scaling and reproduce known implementations such as the expander models~\cite{Ben-Zvi2008,Ben-Zvi2010}. More broadly, understanding how optimization principles at Marr's first level extend to dynamic, growing, mechano-chemical systems and how this predicts and constrains algorithms and mechanisms at the second and third level represents a  major theoretical challenge.

\textbf{Mechanism-function relationships and physical learning}. In some systems, the function directly constrains and shapes the mechanisms. For instance, in the morphogenesis of flow distribution networks, such as plant or animal vasculature, global optima of fluid transport can be achieved by local positive feedback rules coupled with tissue growth~\cite{ronellenfitsch_global_2016}. Similar global optimization through local feedback has been shown to regulate multicellular morphogenesis in slime mold network formation~\cite{alim_random_2013,le_verge-serandour_physarum_2024}. In biophysics, physical local learning has been theoretically shown to provide a route to control the rigidity of tissues~\cite{arzash_rigidity_2025} and cytoskeletal adaptation~\cite{banerjee_learning_2025}. These perspectives connect to the broader concept of physical local learning, in which local and mechanistically realizable feedback rules (Marr's third level) directly achieve global functional optima (Marr's first level)~\cite{stern_supervised_2021,stern_learning_2023}. Mapping out the connections between local learning, the dynamical systems theory, and information-theoretic frameworks is an exciting and timely frontier.

\textbf{Algorithmic origins of pattern information}. While the bounds on information transmission in instructed development are increasingly well understood, we lack predictive theories for how much information can be generated by dynamical rules during self-organization. In principle, any self-organizing patterning system is fully specified by three components: the initial conditions, the boundary conditions, and the underlying dynamical rules, i.e., the algorithms determining the non-local interactions at Marr's second level. Since pure self-organization operates under the rigorous constraint that initial conditions are spatially unpatterned and thus contain no information, the final information is established through amplification and organization of random fluctuations into coherent spatial structures, effectively converting microscopic randomness into macroscopic patterns. The final information content therefore arises from the algorithmic information encoded in the underlying dynamical rules and the geometric constraints imposed by boundary conditions. At the molecular level, these rules and parameters are realized by the sequence and structure of the molecular components, and ultimately encoded in the DNA sequence~\cite{hledik2022accumulation}. For instance, the diffusion coefficient and biochemical interactions of a protein are context-dependent functions of molecular parameters such as the protein's physical size, its sequence and molecular structure. The emergent algorithmic rules are therefore encoded in such ``molecular information.'' This highlights how understanding a given system across Marr's levels may require considering diverse spatial and temporal scales~\cite{tlusty_life_2025}. Resolving how genetic and molecular information determine algorithmic information across scales, and how algorithmic information combines with information encoded in initial and boundary conditions to set the information content of the final output pattern, would therefore tie together concepts across all three of Marr's levels. This represents a crucial challenge for future research.

\section{Acknowledgments}
We thank Edouard Hannezo, Anna Kicheva, Fridtjof Brauns, and all members of the Br\"uckner and Tka\v{c}ik groups for feedback and inspiring discussions.
This work was supported in part by European Research Council ERC-2023-SyG ``Dynatrans'' grant Nr. 101118866 (GT).
This work was conducted while visiting the Okinawa Institute of Science and Technology (OIST) through the Theoretical Sciences Visiting Program (TSVP); at the Kavli Institute for Theoretical Physics (KITP) Santa Barbara, supported by NSF Grant No. PHY-1748958 and the Gordon and Betty Moore Foundation Grant No. 2919.02; and at Lucullus, Vienna.

\bibliography{apssamp}

\end{document}